\documentclass[aps,superscriptaddress,twocolumn]{revtex4-1}
\usepackage{graphicx}
\usepackage{dcolumn}
\usepackage{bm}
\usepackage{colordvi}
\usepackage{color}

\begin{document}

\title{Single spin Landau-Zener-St{\"u}ckelberg-Majorana interferometry of Zeeman-split states with strong spin-orbit interaction in a double quantum dot}
\author{D.V. Khomitsky}
\email{khomitsky@phys.unn.ru}
\affiliation{Department of Physics, National Research Lobachevsky State University of Nizhny Novgorod, 603950 Gagarin Avenue 23, Nizhny Novgorod, Russian Federation}
\author{S.A. Studenikin}
\email{sergei.studenikin@nrc-cnrc.gc.ca} 
\affiliation{Emerging Technologies Division, National Research Council of Canada, Ottawa, ON K1A0R6, Canada}

\begin{abstract}
Single spin state evolution induced by the Landau-Zener-St{\"u}ckelberg-Majorana (LZSM) interference in a Zeeman-spit four level system in a periodically driven double quantum dot is studied  theoretically by the Floquet stroboscopic method. 
An interplay between  spin-conserving and spin-flip tunneling processes with the Electric Dipole Spin Resonance (EDSR) that is induced in an individual dot and enhanced by the LZSM multiple level crossings with the neighboring quantum dot is investigated as a function of the microwave (MW) frequency, driving amplitude, interdot detuning, and magnetic field. 
A number of special points in the parameter space are identified, out of which where all the three features are merged. Under this triple-crossing resonance condition the interdot tunneling is combined with a fast spin evolution in each dot at the EDSR frequency. Harmonics of the EDSR are revealed in the spin-dependent tunneling maps versus variable magnetic field and MW frequency.  The results are applicable for both electron and hole systems with strong spin-orbit interaction and may be useful for developing new time-efficient schemes of the spin control and readout in qubit devices.
\end{abstract}

\date{\today}
\maketitle

\section{Introduction}

The Landau-Zener-St{\"u}ckelberg-Majorana (LZSM) phenomenon manifested in the interference patterns for the transitions between the states \cite{Nori2010,Nori2022,DiGiacomo2005,Nori2018,ShevchenkoNori2018,Ludwig2014,Ludwig2015} is a powerful tool for spectroscopic studies of quantum systems and for manipulation of qubits based on quantum dot (QD) charge \cite{StehlikNori2012,Ludwig2013,Raikh2022}, spin \cite{Rashba,GolovachLoss2006,Koppens2006,Nowack2007,NadjPerge2010,NadjPerge2012,Petersson2012,StehlikPetta2014,StehlikPetta2016,Benito2019,Nori2019,Studenikin2019}, and valley \cite{Petta2018,Burkard2022} degrees of freedom. The rich physics of the multi-level evolution under periodic driving continues to attract interest in studying various structures, including condensed matter \cite{Rudner2008,Schreiber2011,Oosterkamp1998,Granger2015,Studenikin2018,Studenikin2021,Gomez2019,Grimaudo2020,Kitamura2020,Chen2021,Malla2021,Zhou2014,Pasek2018}, { interacting Josephson qubits \cite{SataninNori2012,SataninNori2014,Bastrakova2021}} and atomic systems \cite{DiGiacomo2005,Vasilev2007,Novelli2015,Liang2020,Liu2021,Nath2020,Nath2021}.
Thanks to the strong spin-orbit interaction (SOI), for example, observed in hole spin devices \cite{Studenikin2018,Studenikin2019} and narrow band-gap semiconductors \cite{NadjPerge2010,NadjPerge2012,Petersson2012,StehlikPetta2014,StehlikPetta2016}, the spin levels become strongly coupled during the LZSM level-crossing processes. This leads to new spin-dependent phenomena and new opportunities for research and applications. Additional interest to hole spins is driven by the prediction of reduced noise caused by the hyperfine interaction with the nuclei spins \cite{Coish2008,Fallahi2010}.

In the present work, along with LZSM interference, we bring one more phenomenon into play, the electric dipole spin resonance (EDSR) \cite{Rashba,GolovachLoss2006,Koppens2006,Nowack2007,NadjPerge2010,NadjPerge2012,Ludwig2015b}. In particular, we study LZSM-induced spin-dependent tunneling and single-spin evolution in a periodically driven system of spin levels with strong spin-orbit coupling.
Being the cause for the inter-dot spin flip tunneling, the SOI is also responsible for another important phenomenon - the EDSR that lies at the basis of the spin operations for quantum information applications. In Ref.\cite{Studenikin2019} an efficient SOI was achieved by employing strong interdot coupling on the order 100 $\mu eV$ between the Zeeman-split spin levels in the neighboring dots. Such a strong coupling leads to a noticeable admixture of spin states, resulting in a finite individual spin-flip transition probability under the EDSR conditions. In the present work we explore a situation with much weaker interdot coupling on the order 1 $\mu eV$, which is close to the experimental conditions in Ref.\cite{Studenikin2018}.    

In this paper we use the Floquet stroboscopic approach to explore evolution of Zeeman-split states in a GaAs-based double quantum dot (DQD) with a multi-level structure and carefully examine the hybrid situation when LZSM and EDSR processes occur at the same time. The LZSM transitions in a multilevel system have been considered before for the linear \cite{Sinitsyn2015} or perturbative and in general a nonlinear \cite{Ashhab2022} approximation of time dependence for the field-driven levels. In our model we consider a periodic driving field which naturally requires the application of the Floquet stroboscopic technique.
This approach can be successfully applied for the description of both the tunneling and spin evolution if both processes are triggered on a time scale of many driving periods  \cite{GrifoniHanggi1998,Hanggi1991,Khomitsky2012,Qiao2021}, which is the case of our model.  All the three types of transitions described above (spin-conserving tunneling, spin-flip tunneling and the EDSR) are revealed in our simulations. As a test, for a simple tunneling regime we apply the well-known analytical expressions for two-level LZSM patterns \cite{Nori2010,Nori2018,Nori2022} and find their good agreement with our numerical simulations in the framework of the Floquet approach. We examine various points in the parameter space and also find the conditions under which all the three transitions take place simultaneously. Such a hybrid resonance cannot be described within a two-level model. At minimum a four-level model is required to simulate the spin evolution of the system under study. This regime includes the fast interdot tunneling with the spin flip where the spin flip is observed in both dots of the DQD system. The observed spin dynamics under the hybrid resonance has a rather complex pattern which resembles the evolution under the time-shaped profile of the electric field pulses aimed at speeding up the spin flip time \cite{Budagosky2016}.  
The advantage of tuning the system into such a hybrid regime is a much faster spin flip  transitions in the same QD where the spin state has been initialized. We predict an enhanced precession by several times faster  compared to the known spin-orbit induced EDSR mechanism in an individual dot \cite{Nowack2007}.  
Another important difference of our four-level model with the two-level EDSR mechanism \cite{Nowack2007} is a notable and tunable enhancement of the spin-flip Rabi frequency observed as a non-linear function of the driving strength due to the tunnel coupled states in the neighboring QD \cite{Khomitsky2012}. 

The main advances of the present manuscript from the line of the results achieved in Ref \cite{Studenikin2018} by which our study is largely motivated are the following ones. First, we derive the primary matrix elements defining the typical timescales of various processes (spin-conserving tunneling, spin-flip tunneling, EDSR) directly by the first principles from the eigenstates in a realistic double dot potential profile reflecting the actual configurations achieved in experiments.  
Second, we explore the space of the system parameters in a greater variety of directions where different combinations of the parameters are considered compared to Ref \cite{Studenikin2018}. In particular, our main findings can be seen most clearly in a newly considered Magnetic field - Driving frequency plane.  Third, we explore the quantum state dynamics for the selected points in the parameter space in the time domain both for the level occupations, charge and spin average values that helps in understanding the entangled dynamics of charge and spin. Finally, when possible our numerical results are compared with the well-established analytical models of the LZSM interferometry from Refs \cite{Nori2010,Nori2022}.

This paper is organized as follows. In Sec. II we introduce the Hamiltonian of our system and discuss the time-independent and periodic contributions as well as the observables. In Sec. III we discuss the primary regimes of the evolution in terms of the associated resonance conditions and within the framework of a two-level analytical model. { We focus on the spin-conserving tunneling, the spin-flip tunneling and the EDSR as well as on the regime where all the three modes are merged together. It turns out that the onset of each of the regimes can be described by simple analytical conditions.} In Sec. IV  we introduce the numerical parameters and describe the tunneling and spin dynamics on the maps of averaged total and spin-dependent tunneling probability. In Sec. V we present the evolution examples for all principal regimes of our system in terms of the dynamics of observables and level populations. Finally, in Sec. VI we summarize the results.

\section{Model and observables}

Our model is based on the solution of the non-stationary Schr{\"o}dinger equation with the Hamiltonian typical for the 1D models of double quantum dots with spin-orbit coupling and subject to constant magnetic and periodic electric fields \cite{Khomitsky2012}:

\begin{equation}
H=H_{\rm{2QD}}+H_Z+H_{\rm{SO}}+V(x,t).
\label{ham}
\end{equation}

This Hamiltonian describes the dynamics and the tunneling which take place essentially in 1D channel connecting the QDs. This is close to the situation realised in recent experiments \cite{Studenikin2018} were a DQD was created in a two-dimensional hole gas by the surface electrostatic gates and the tunneling occurred in one dimension along the line connecting the dots. The tunneling is described in a single-particle approximation, although the models of double dots with two electrons or holes working as two-qubit systems with time-dependent control have also attracted a considerable attention \cite{Taylor2007,Platero2022}. 
In (\ref{ham}) $H_{\rm{2QD}}=k_x^2/2m+U_0((x/d)^4-2(x/d)^2)$ is the Hamiltonian of the hole with the effective mass $m$ in the lowest subband of size quantization with the direction of the Ox axis pointing along the double dot structure (hereafter we use units with $\hbar=1$). Here the symmetric double well potential is described by the interdot center distance $2d$ and the barrier height $U_0$. The next term in (\ref{ham}) is the Zeeman coupling term: 

\begin{equation}
H_Z=\frac{1}{2}g\mu_B B_z \sigma_z 
\label{hz}
\end{equation}

produced by the constant magnetic field along Oz axis and g is the effective hole g-factor, the parameter controlling the  Zeeman splitting energy and the EDSR condition \cite{Studenikin2018,Studenikin2019}:

\begin{equation}
\Delta_Z=g\mu_B B_z. 
\label{deltaz}
\end{equation}

\begin{figure}[tbp]
\centering
\includegraphics*[width=0.5\textwidth]{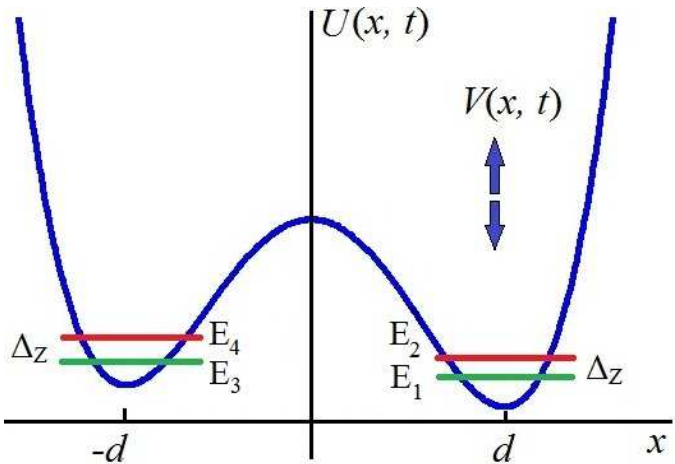}
\caption{Scheme of the potential profile $U(x,t)=H_{\rm{2QD}}+V(x,t)$ described by Hamiltonian (\ref{ham}) with double dot potential $H_{\rm{2QD}}$ altered by detuning and driving $V(x,t)$. The individual dot potential minima are at $\pm d$ and the spin-resolved ground doublet E1, E2 in the right QD and E3, E4 in the left QD are characterized by the Zeeman splitting $\Delta_Z$.}
\label{figdqd}
\end{figure}

The third term in Eq.(\ref{ham}) can be expressed as following: 

\begin{equation}
H_{\rm{SO}}=\beta_D \sigma_x k_x 
\label{hso}
\end{equation}

where $\beta_D$ is the strength of the bulk Dresselhaus SOI which contains the contribution linear in the wavevector being the leading terms for  GaAs-based low-dimensional structures \cite{GolovachLoss2006}.  
The last term $V(x,t)$ describes the static detuning and driving produced by the periodic electric field. For $t<0$ only the static detuning potential is present applied mainly to the right QD,

\begin{equation}
V(x,t<0)=U_d f_d(x),
\label{vxtl0}
\end{equation}

where $f_d=(x/d_1)^3-3/2\cdot (x/d_1)^2$ models the smooth fitting with the initial double well potential with $d_1=1.5 d$. For $t \ge 0$ the additional periodic driving is turned on,

\begin{equation}
V(x,t \ge 0)=\left[U_d+V_d \sin \omega t \right]f_d(x).
\label{vxtg0}
\end{equation}

In (\ref{vxtl0}) and (\ref{vxtg0}) $U_d$ is the detuning amplitude with $U_d<0$ corresponding to the right QD bottom shifted down and $V_d$ is the driving amplitude. The sum of the double well potential $H_{\rm{2QD}}$ and the detuning (\ref{vxtl0}) creates the potential profile $U(x,t)=H_{\rm{2QD}}+V(x,t)$ sketched on Fig.\ref{figdqd} showing the double well with two pairs of Zeeman-split levels located in each of the two quantum dots.  
We perform the numerical diagonalization of the time-independent part of the Hamiltonian (\ref{ham}) for a multilevel double QD and obtain the set of energy levels $E_n$ and the eigenfunctions $\phi_n(x)$, the latter being two-component spinors. We build a multilevel ensemble of states representing the actual basis with the degree of completeness required for the calculation of all the matrix elements for finding the tunneling and spin flip transition probabilities.

The solution for the time-dependent Schr{\"o}dinger equation is found as a sum of the eigenfunctions of the stationary part of the Hamiltonian with time-dependent coefficients:

\begin{equation}
\psi(x,t)=\sum_n C_n(t) e^{-i E_n t} \phi_n(x).
\label{psixt}
\end{equation}

In the present work during the construction of the wavefunction (\ref{psixt}) we restrict ourselves to the subspace of four lowest levels $E_1, \ldots, E_4$ depicted in Fig.\ref{figdqd} since the primary regimes of the evolution observed in the experiments \cite{Studenikin2018,Studenikin2021} can be described in a four-level approximation. In each QD we consider a ground pair of spin-resolved levels $E_1$, $E_2$ and $E_3$, $E_4$, respectively, with primarily opposite z-projections of spin in the presence of the Zeeman term ({\ref{hz}}) depicted by the green and red line colors in Fig.\ref{figdqd}. Our full modeling demonstrates that around 93\% of the wavefunction norm is contained within the 4-level subspace throughout the whole dynamics. These four levels correspond to the lowest levels of the system representing the charge and spin degrees of freedom, i.e. the states localized in the left or right QD, respectively, each having the spin up or spin down, similar to the model that has been adopted in \cite{Studenikin2018}. In the present paper we explore the evolution in a much wider area of the parameter space compared to \cite{Studenikin2018} discovering various regimes described in the next Section. The set of ordinary differential equations are obtained for the coefficients $C_n(t)$ in (\ref{psixt}) which depend on the matrix elements $V_{nl}$ of the driving potential (\ref{vxtg0}). This system is accompanied by the initial condition $C_n(0)$ describing the spin-down wavepacket injected into the right QD which replicates the experimental conditions in \cite{Studenikin2018,Studenikin2021}. The described system is solved via the standard Cayley numerical scheme in the Floquet stroboscopic representation where the results are presented at the time moments measured in units of the driving period, $t=NT$ where $T=2\pi / \omega$ is the driving field period. The observables are calculated using the reconstructed wavefunction (\ref{psixt}) across the whole double QD at the given moment of stroboscopic time $t=NT$. The first observable is the time-dependent probability $P_{L}(t)$ to find the particle in the left QD representing the tunneling efficiency defined as

\begin{equation}
P_{L}(t)=\int_{-\infty}^{0} \mid \psi(x,t) \mid^2 dx.
\label{plr}
\end{equation}

The electrical current through the DQD under consideration \cite{Studenikin2018} is produced by holes tunneling from the right lead, between the dots, and to the left lead as shown in Fig.\ref{figlevels}a. Therefore, the electrical current through the DQD system is proportional to the $P_L$  averaged over the observation time. The second observable is the probability $P_{R}(t)$ to find the particle in the right QD which is found from the normalization condition $P_{L}(t)+P_{R}(t)=1$. Since in our model the spin enters as another degree of freedom we will be interested also in calculating two more observables: the z-projection of spin $\sigma_z^{(L,R)}(t)$ measured in left or right QD, respectively:

\begin{equation}
\sigma_z^{(L)}(t)=\int_{-\infty}^{0} \langle \psi |\sigma_z| \psi \rangle dx,
\label{sigmazl}
\end{equation}

\begin{equation}
\sigma_z^{(R)}(t)=\int_{0}^{\infty} \langle \psi |\sigma_z| \psi \rangle dx.
\label{sigmazr}
\end{equation}

We calculate the time-averaged tunnel probability and the spin-dependent tunnel probabilities for the $N$ driving periods as following:

\begin{equation}
P_{L}=\frac{1}{NT} \int_0^{NT} P_{L}(t) dt,
\label{psaver}
\end{equation}

\begin{equation}
\sigma_z^{(L,R)}=\frac{1}{NT} \int_0^{NT} \sigma_z^{(L,R)}(t) dt.
\label{szaver}
\end{equation}

The number $N$ of the driving periods used to obtain the maps of the averaged values (\ref{psaver}), (\ref{szaver}) depends on the typical timescales of the evolution on which the steady picture is formed. These timescales primarily depend on the matrix elements of typical transitions in the system. Our explicit calculations of these matrix elements allowed us to limit the evolution stroboscopic time to $N=500 \ldots 1000$ driving periods for most of the regimes considered. These limits agree in general with the ones used in the experiments \cite{Studenikin2018}.
It should be also noted that due to the presence of SOI term (\ref{hso}) the spin is in general no longer conserved during the evolution in the whole space of the system states, and the contributions (\ref{sigmazl}), (\ref{sigmazr}) are not coupled via the normalization condition. Nevertheless, we use these observables to visualize the spin evolution during the LZSM process in both dots since the SOI may trigger spin flips both with and without the tunneling, as we will discuss below.

\section{Regimes of evolution and the two-level approximation}

\begin{figure}[tbp]
\centering
\includegraphics*[width=0.5\textwidth]{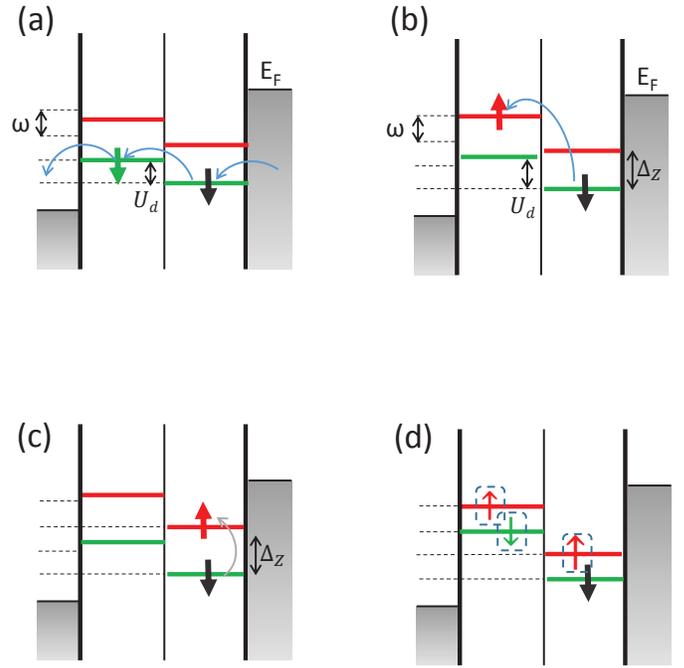}
\caption{Four-level system with the Zeeman doublet splitted by $\Delta_Z$ in left and right QDs which bottoms are shifted by the detuning $U_d$. The initial state (black arrow) is the spin-down state in the right QD and the final state is marked by the green or red arrow. The principal transitions triggered by the driving field include (a) spin-conserving tunneling during the LZSM level passage; (b) spin-flip tunneling during the same passage; (c) EDSR in the right QD without tunneling; (d) hybrid regime with LZSM level passage and the EDSR spin flip where the final state labeled by framed arrows has contributions from all four levels.} 
\label{figlevels}
\end{figure}

Different kind of transitions can occur in the four-level system shown in Fig.\ref{figdqd}. In Fig.\ref{figlevels} we show schematically the level structure and the basic regimes of the evolution which can be triggered by the periodic electric field with the potential $V(x,t)$. 
Its discrete resonance action can be formally described as due to resonances with certain number of photon quanta $k \omega$, the phenomenon commonly referred as the Photon Assisted Tunneling (PAT) \cite{Nori2010,Studenikin2018,Studenikin2021,GrifoniHanggi1998,Zhang2006}. 
It should be noted that the basic mechanism behind it being the interference created by the multiple level passage during the periodic driving, i.e. of the LZSM-type. We will still continue to call such a situation as the PAT-regime due to the discreet character of the pattern in the energy space. The initial and final states assigned to the spin-up or spin-down level in the corresponding QD are labeled in Fig.\ref{figlevels} by the black and green or red arrows, respectively, indicating the spin projection.

The following regimes of driven evolution can be identified in Fig.\ref{figlevels}.
Figure\ref{figlevels}a depicts the spin-conserving tunneling when a number of photon quanta equals the detuning amplitude $|U_d|$ that corresponds to the PAT regime:

\begin{equation}
|U_d|=k_1 \omega.
\label{pat} 
\end{equation}

Figure\ref{figlevels}b shows the spin-flip tunneling when a number of photon quanta equals the detuning amplitude plus the Zeeman splitting providing the hole to tunnel to the level with another spin projection,

\begin{equation}
|U_d|+\Delta_Z=k_2 \omega.
\label{patflip} 
\end{equation}

Figure\ref{figlevels}c presents the EDSR without tunneling that takes place in a single QD when the following condition is satisfied:

\begin{equation}
\Delta_Z=k_3 \omega
\label{edsrcond}
\end{equation}

i.e. the driving frequency itself ($k_3=1$) or one of its harmonics ($k_3=2,3,...$) matches the Zeeman splitting (\ref{deltaz}) calculated in the presence of SOI.

The specific feature of the regimes with resonances (\ref{pat}) - (\ref{edsrcond}) is that they involve basically a pair of two levels in the dynamics. The tunneling regimes (\ref{pat}),(\ref{patflip}) can be described by the driving applied mainly to one level only located in the right QD while in the EDSR regime both levels in right QD are driven with the same profile in time domain. Thus, the former cases (\ref{pat}),(\ref{patflip}) fall within the limits of the well-known two-level driven model where the distance between the levels has a constant detuning part $U_d$ plus the periodic modulation $V_d \sin \omega t$ \cite{Nori2010,Nori2022}. To determine whether a slow or fast limit of the evolution is present one needs to estimate the adiabaticity parameter \cite{Nori2010,Nori2022,DiGiacomo2005}

\begin{equation}
\delta=\frac{\Delta^2}{4 v}
\label{adiabpar}
\end{equation}

where $\Delta$ is the tunneling amplitude (level coupling) and $v=d(E_2-E_1)/dt$ is the rate of level distance change in energy space which can be estimated for periodic driving as $v=\omega V_d$. For the system under consideration \cite{Studenikin2018} typical values are $\Delta \sim 1$ $\mu eV$, $\omega \sim 10$ $\mu eV$, $V_d \sim 100$ $\mu eV$ so the adiabaticity parameter $\delta \sim 10^{-3} ... 10^{-4}$ which indicates the fast-passage limit. In such limit the averaged probability $P$ of a transitions in a two-level system is given by \cite{Nori2010,Nori2018,Nori2022}

\begin{equation}
P=\frac{1}{2} \sum_k \frac{\Delta_k^2}{(k \omega-U_d)^2+\Delta_k^2}
\label{ptwolev}
\end{equation}

where $U_d$ is the detuning and $\Delta_k=\Delta J_k(V_d/ \omega)$ where $J_k$ is the k-th Bessel function. In the next Sec. we will apply the analytical estimate (\ref{ptwolev}) for comparison with our numerical results for the tunneling probability.

It should be mentioned that all of the types of evolution regimes with the resonances (a) - (c) are known for driven dynamics in double dots \cite{Studenikin2018,Studenikin2021,Khomitsky2012}. In our model a new  hybrid regime is identified  which scheme is depicted on panel (d) in Fig.\ref{figlevels}. Here all the conditions (\ref{pat}) - (\ref{edsrcond}) are satisfied simultaneously which takes place when

\begin{equation}
k_2=k_1+k_3.
\label{hybrcond}
\end{equation}

The level scheme shown in Fig.\ref{figlevels}(d) has the triple framed green arrow indicating that in a hybrid regime one can find the final state with certain probability on all three states besides the initial one. From the dynamical point of view such hybrid regime means that we may observe the partial spin-conserving and spin-flip tunneling happening on the EDSR frequency or on its harmonics. 
It should be noted that the hybrid resonance (\ref{hybrcond}) has a universal character because the condition (\ref{hybrcond}) can be satisfied for any set of system parameters which includes the double dot potential, the detuning and the driving field strength. The only two parameters which have to be adjusted is the driving frequency and the magnetic field. First, we fix the driving frequency in accordance with the spin-conserving tunneling condition (\ref{pat}) where any integer $k_1$ can be chosen. Second, with the defined frequency we fix the magnetic field in accordance with the EDSR condition (\ref{edsrcond}) again choosing any integer $k_3$. After that one can see that the spin-flip tunneling is also possible since the condition (\ref{patflip}) is also fulfilled if one chooses the $k_2$ integer to satisfy the condition (\ref{hybrcond}). From the practical point of view it means that the hybrid resonance points can always be found in the map of (Driving frequency, Magnetic field) parameters which will be considered in the following Section. 

\begin{figure*}[tbp]
\centering
\includegraphics*[width=0.95\textwidth]{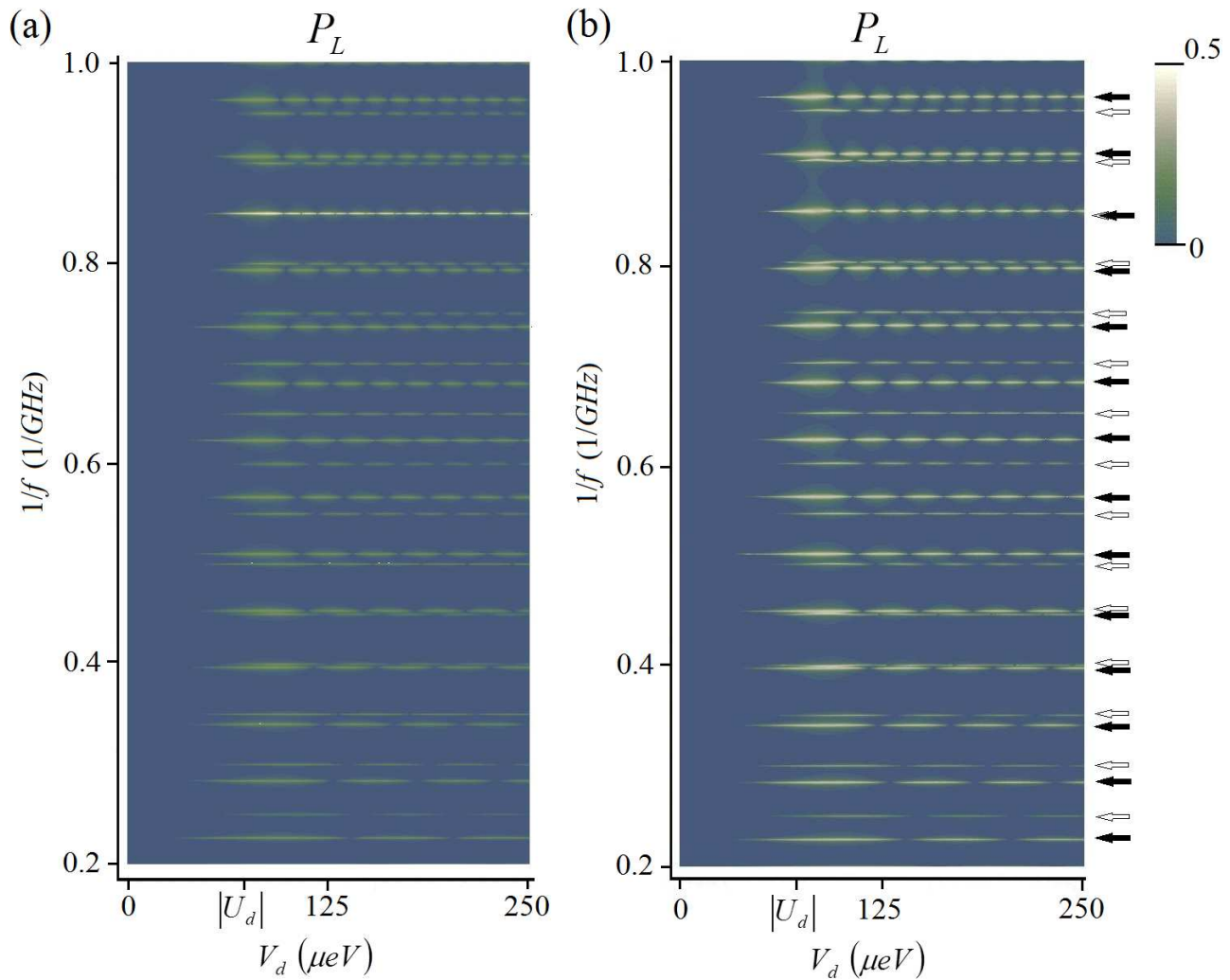}
\caption{Contour maps of interdot tunneling probability $P_L$ (\ref{psaver}) calculated by (a) analytical two-level approach (\ref{ptwolev}) and (b) found numerically by (\ref{psaver}). Filled arrows mark the spin-conserved tunneling (\ref{pat}) and open arrows mark the spin-flip tunneling (\ref{patflip}). The tunneling becomes effective when the driving exceeds the detuning at $V_{d}>|U_d|$.}
\label{FigVdrinvf}
\end{figure*}

Finally, for the off-resonance driving field where none of the conditions (\ref{pat}) - (\ref{edsrcond}) is satisfied the driving does not produce any change of the initial state describing the spin-down particle in the right QD. The localization in a single dot of a double dot system during the application of a periodic driving with certain amplitude and frequency for the case of zero detuning is known as the coherent destruction of tunneling (CDT) \cite{GrifoniHanggi1998,Hanggi1991}. Here the destructive interference expressed via the quasienergy crossing can brought the tunneling to a standstill. Similar localization in the non-resonant 2D areas on tunneling probability maps plotted for a pair of system parameters can be observed for our system with finite detuning $U_d$ even for large driving amplitudes $V_d>|U_d|$ as it will be seen below in the right-hand parts of the panels in Fig.\ref{FigVdrinvf} and Fig.\ref{FigVdrinvf2}. The existence of such areas of largely suppressed tunneling indicates that for a finite detuning a resonance condition like (\ref{pat}) or (\ref{patflip}) is required for the tunneling to be effective.

\section{Numerical parameters and tunneling probability maps}

 Let us proceed with the numerical results obtained for the hole GaAs DQD structure with parameters similar to those in \cite{Studenikin2018}: the hole effective mass $m_h=0.11 m_0$, the interdot minima distance $2d=116$ nm, the barrier height $U_0=5$ meV, the g-factor $g=1.35$ and the SOI Dresselhaus constant $\beta_D=3$ $\rm{meV} \cdot \rm{nm}$. Our calculation show that  under this conditions the spin-conserving tunneling rate is about 1 $\mu eV$, and the spin-flip tunneling rate is about 0.45 $\mu eV$, that is close to the experiments in \cite{Studenikin2018}. The initial state is the spin-down wavepacket with width $\sim d$ centered in the right QD represents the hole injected from the right lead to the ground state of the right QD in accordance with the experimental settings in Ref.\cite{Studenikin2018}. In our study we employ two-dimensional (2D) maps of averaged tunneling probability (\ref{psaver}) and spin-dependent tunneling probability (\ref{szaver}) in the plane of specifically chosen sets of parameters where the different  regimes depicted in Fig. 2 are identified and explored.

\subsection{Tunneling in the plane of driving amplitude and inverse  frequency}   

We start with the building of 2D maps for averaged probabilities (\ref{psaver}), (\ref{szaver}) under the fixed magnetic field $Bz=0.125$ T and fixed detuning $U_d=-73$ $\mu eV$ corresponding to the ground state being a spin-down state in the right QD at the Zeeman splitting $\Delta_Z=9.75$ $\mu eV$. The two variable parameters are the driving amplitude $V_d$ and frequency $\omega=2\pi f$. Since the values $k_j$ of resonance maxima in (\ref{pat}) - (\ref{edsrcond}) are inversely proportional to the frequency $\omega$ it is more convenient to plot the maps in coordinates $(V_d,1/f)$. First, in Fig.\ref{FigVdrinvf}a we show the contour plots for tunneling probability obtained from the analytical estimate (\ref{ptwolev}) in the amplitude range $V_d=0...250$ $\mu eV$ and the frequency band $f=1...5$ GHz for two sets of the PAT transitions with resonances given by (\ref{pat}) marked by the solid arrows and by (\ref{patflip}) marked by the open arrows being the spin-conserving and spin-flip tunneling transitions, respectively. In Fig.\ref{FigVdrinvf}b we show the corresponding map obtained numerically from (\ref{psaver}). A very good agreement is obvious between the two maps confirming that the numerical procedure is correct and therefore can be used for more complex situations where all four levels play role in the spin state evolution. Note that the tunneling becomes effective when the driving amplitude $V_{d}$ exceeds the interdot detuning amplitude $|U_d|$, i.e. to the right of the line $V_{d}=|U_d|$.

\begin{figure*}[tbp]
\centering
\includegraphics*[width=0.95\textwidth]{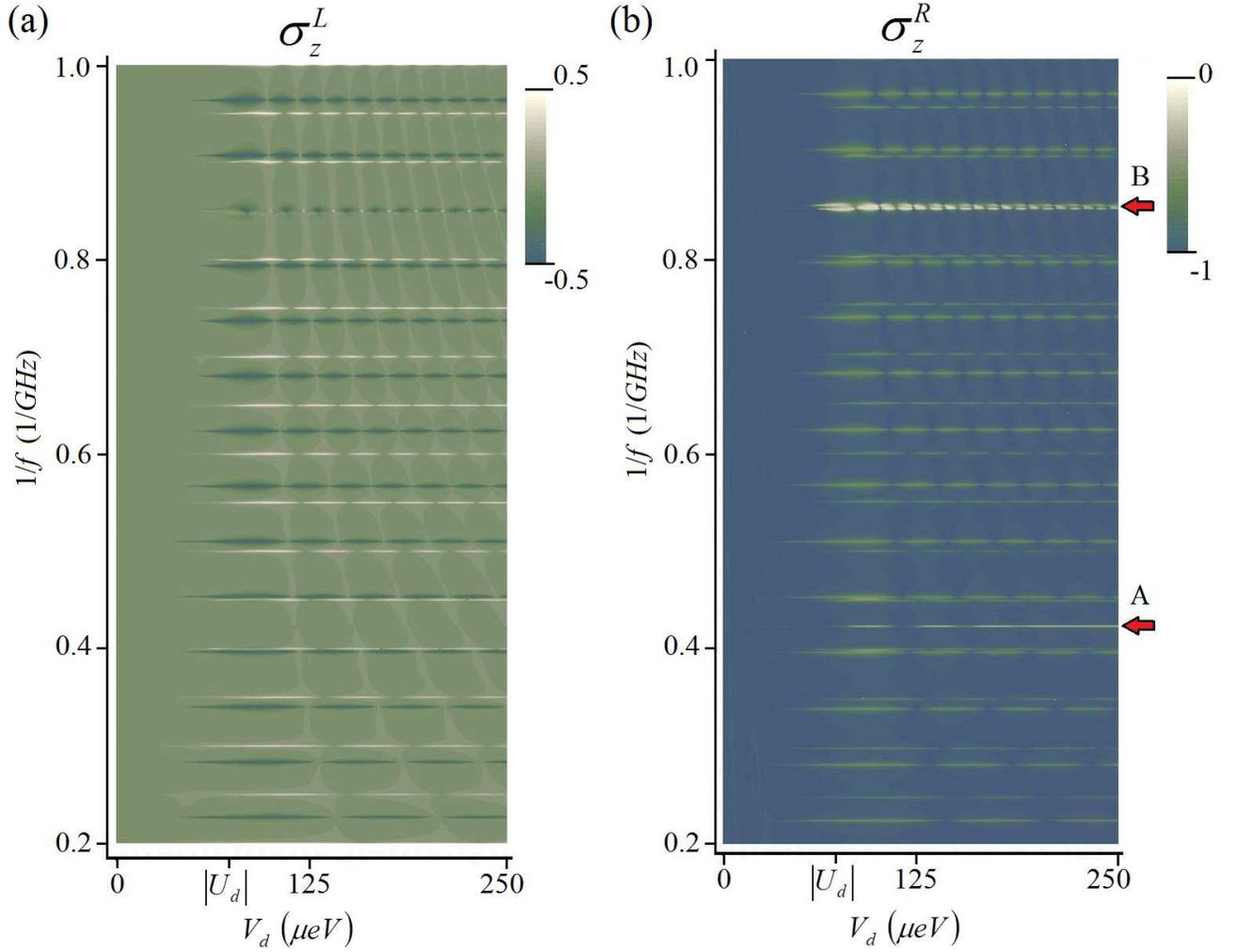}
\caption{Maps of spin-dependent tunneling probability (\ref{szaver}) in (a) left QD and (b) right QD corresponding to Fig.\ref{FigVdrinvf}. In panel (a) dark lines correspond to the spin-conserving tunneling and bright lines correspond to the spin-flip tunneling. In panel (b) both types of tunneling are shown by bright lines. The additional EDSR lines in panel (b) described by (\ref{edsrcond}) are marked by arrows. Arrow A marks the primary EDSR line $k_3=1$ and arrow B marks its second harmonic $k_3=2$.}
\label{FigVdrinvf2}
\end{figure*}

To explore the spin-dependent tunneling let us examine  the spin profile of the tunneling and plot the maps of the averaged spin projection (\ref{szaver}) in left and right QDs. In Fig.\ref{FigVdrinvf2} we show such maps for the same parameters as in Fig.\ref{FigVdrinvf}. It should be mentioned that while the observables (\ref{plr}) and (\ref{sigmazl}), (\ref{sigmazr}) vary in the intervals $(0,1)$ and $(-1,1)$, respectively, their time averages (\ref{psaver}) and (\ref{szaver}) have in general lower bounds which explains the different limiting values of color bars in the 2D maps discussed here. 
In Fig.\ref{FigVdrinvf2} one can see that the spin tunneling maps are described by two sets of maxima lines. In Fig.\ref{FigVdrinvf2}a the first set of lines is dark corresponding to  the spin-conserving tunneling where the negative spin projection is maintained and follows the resonance condition (\ref{pat}). The second set of lines is bright corresponding to the spin-flip condition (\ref{patflip}) where the spin projection is flipped to positive values during the tunneling. Both corresponding sets of maxima lines are bright in panel (b) for the right QD since any type of tunneling lifts the averaged spin projection from the background dark color corresponding to the value $\sigma_z=-1$. It should be noted that the conditions (\ref{pat}) and (\ref{patflip}) may provide the close frequencies for certain combinations of parameters and values of $k_j$. This means that some of the lines from different families can, in principle, be very close to each other. By examining both panels of Fig.\ref{FigVdrinvf2} one may notice that it indeed happens for the values $k_1=15$ in (\ref{pat}) and $k_2=17$ in (\ref{patflip}) marked by the arrow B described below. Fig.\ref{FigVdrinvf2}b shows the averaged spin $\sigma_z^{(R)}$ in the right QD. In the lower part of Fig.\ref{FigVdrinvf2}b we see the EDSR line marked by the arrow A corresponding to the main EDSR line $k_3=1$ in (\ref{edsrcond}) and in the upper part we see its second harmonic corresponding to $k_3=2$ in (\ref{edsrcond}) marked by arrow B. For the chosen parameters the EDSR harmonic marked by arrow B is very close to the tunneling lines $k_1=15$ and $k_2=17$ discussed above. This is an example of the situation when all the three resonances (\ref{pat}) -  (\ref{edsrcond}) coincide forming a special hybrid resonance with the four levels involved. Such situations when the three resonance lines cross can be more conveniently revealed in the map of variables $(Bz, f)$ with the variable magnetic field and the driving frequency which will be considered in the following Subsection.

\begin{figure*}[tbp]
\centering
\includegraphics*[width=0.9\textwidth]{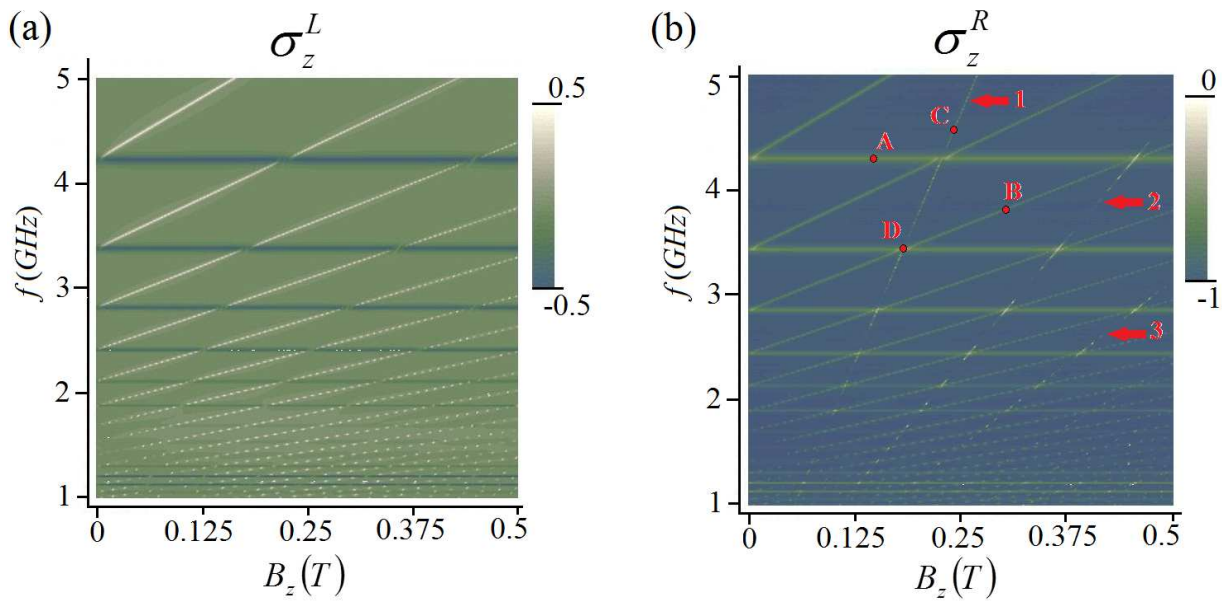}
\caption{Maps of time-averaged spin-resolved tunneling probability (\ref{szaver}) in (a) left QD and (b) right QD shown in $(Bz, f)$ plane at fixed detuning $U_d=-70$ $\mu eV$ and fixed driving amplitude $V_d=100$ $\mu eV$. Dark horizontal lines in panel (a) correspond to the spin-conserving tunneling satisfying the condition (\ref{pat}). Bright angled lines in panel (a) are for spin-flip tunneling satisfying the condition (\ref{patflip}). In the right QD on panel (b) additional steeper angled lines marked by arrows $1,2,3$ represent the EDSR lines satisfying (\ref{edsrcond}) with harmonics $k_3=1,2,3$. The crossing of all three lines in points like point D can happen under the condition (\ref{hybrcond}) of the hybrid resonance. Evolution for selected points A...D in the time domain is discussed in the text.}
\label{figbzf}
\end{figure*}

\subsection{Probability maps in the 2D plane of magnetic field and driving frequency}

Since the main tunneling features can be captured by the spin-resolved tunneling probability maps in this Subsection, we focus on them only. We set the detuning $U_d=-70$ $\mu eV$ for the ground state as the spin-down state in right QD and fix the driving strength $V_d=100$ $\mu eV$. The magnetic field is varied between 0 and 0.5 T, and the frequency is varied in the same band 1...5 GHz.  In Fig.\ref{figbzf} we show the contour plots of the averaged spin projection (\ref{szaver}) in (a) left QD and (b) right QD, respectively. The spin-conserving tunneling is independent of the magnetic field and is expressed via dark horizontal lines in panel (a) and bright horizontal lines in panel (b), each following the resonance condition (\ref{pat}). An example of such tunneling represented by the point A in panel (b) will be discussed in the next Section in the time domain. The spin-flip tunneling is magnetic field-dependent and is expressed via angled bright lines in both panels following the resonance condition (\ref{patflip}). An example of such tunneling marked by point B will be discussed in the next Section. Finally, in panel (b) there is another family of steeper angled lines corresponding to the EDSR in the right QD which is described by the resonance condition (\ref{edsrcond}). The lines marked by arrows $1,2,3$ represent the EDSR lines satisfying (\ref{edsrcond}) with harmonics $k_3=1,2,3$. An example of such evolution marked by point C will be discussed below. 

One can see that at certain points of the $(Bz, f)$ plane in Fig.\ref{figbzf}b the resonance lines belonging to all three families (\ref{pat}) - (\ref{edsrcond}) cross each other. It happens when the condition (\ref{hybrcond}) is satisfied at points such as point D marked by red circle in Fig.\ref{figbzf}b. These points represent the hybrid resonance which cannot be described in terms of two-level system, as we will see in the next Section by considering the evolution for selected points in the time domain.

\section{Time evolution of the observables and level occupations}

\begin{figure*}[tbp]
\centering
\includegraphics*[width=0.95\textwidth]{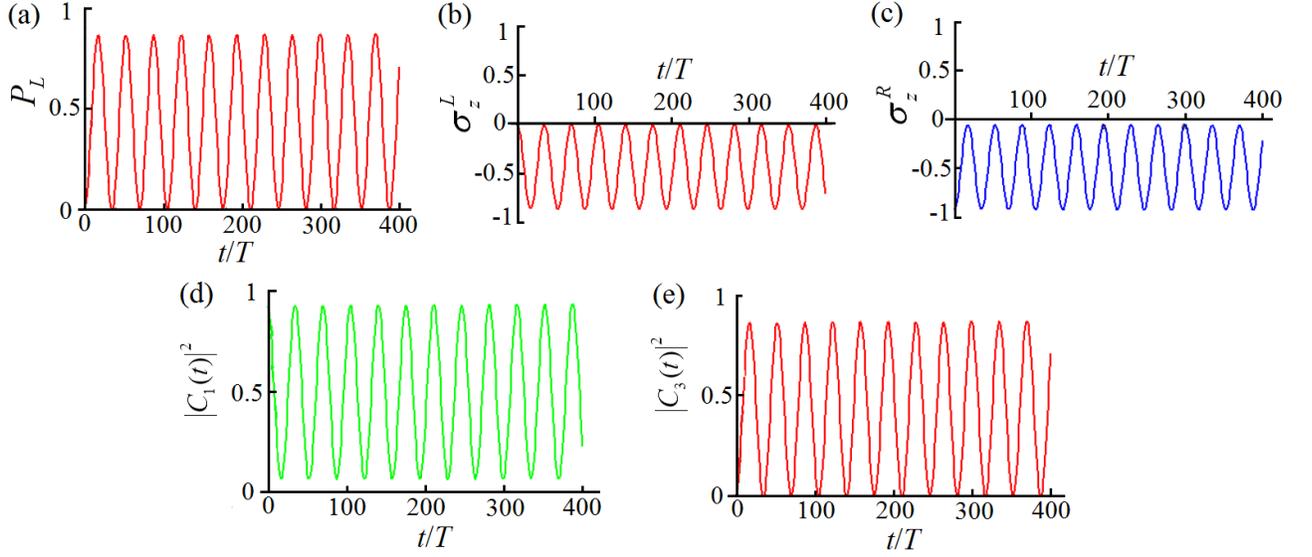}
\caption{Time dependence on 400 driving periods for the point A in Fig.\ref{figbzf}(b) showing (a) tunneling probability and (b), (c) spin projection in left and right QD. The spin-conserving tunneling is observed. (d), (e) Evolution of the occupation probabilities for the states $E_1$, $E_3$ from Fig.\ref{figdqd} participating in the dynamics which has essentially two-level character.}
\label{figa}
\end{figure*}

Here we turn our attention to the stroboscopic evolution  of of the observables and the level occupation probabilities for the selected points A, B, C, D in Fig.\ref{figbzf}(b) to be shown in time domain at $t=NT$ where $T$ is the driving field period. We start with point A located on the PAT line (\ref{pat}) with $k_1=4$ and representing the spin-conserved tunneling. The evolution of selected observables and level populations is shown in Fig.\ref{figa}. Here the time-dependent tunneling probability $P_L$ (\ref{plr}) in panel (a) exhibits oscillations with the period $2\tau_t$ where $\tau_t  \sim 17 T$ is the tunneling time. For the given frequency we have $\tau_t \sim 5$ ns which inverse corresponds to the typical values of spin-conserving matrix element $\Delta_c \sim 1$ $\mu eV$ coupling the states of the same spin in left and right QD. The spin projections (\ref{sigmazl}) and (\ref{sigmazr}) shown in panels (b) and (c) reproduce the spin-down population in left and right QD following the tunneling probability behavior. They demonstrate oscillations with the same period and with the same average negative value indicating the conservation of spin during the tunneling in the PAT regime. In the following panels (d), (e) we show the evolution of the populations for the states $E_1$ and $E_3$ from Fig.\ref{figdqd} which are the only states essentially involved in the dynamics for the point A. These states are the spin-down states in the right and left QDs, respectively, and their oscillating population reflects the spin-conserving tunneling described above. We see that here the dynamics can be described in the framework of the two-level subspace where the transition probability is given by (\ref{ptwolev}) under the resonance condition (\ref{pat}).

\begin{figure*}[tbp]
\centering
\includegraphics*[width=0.95\textwidth]{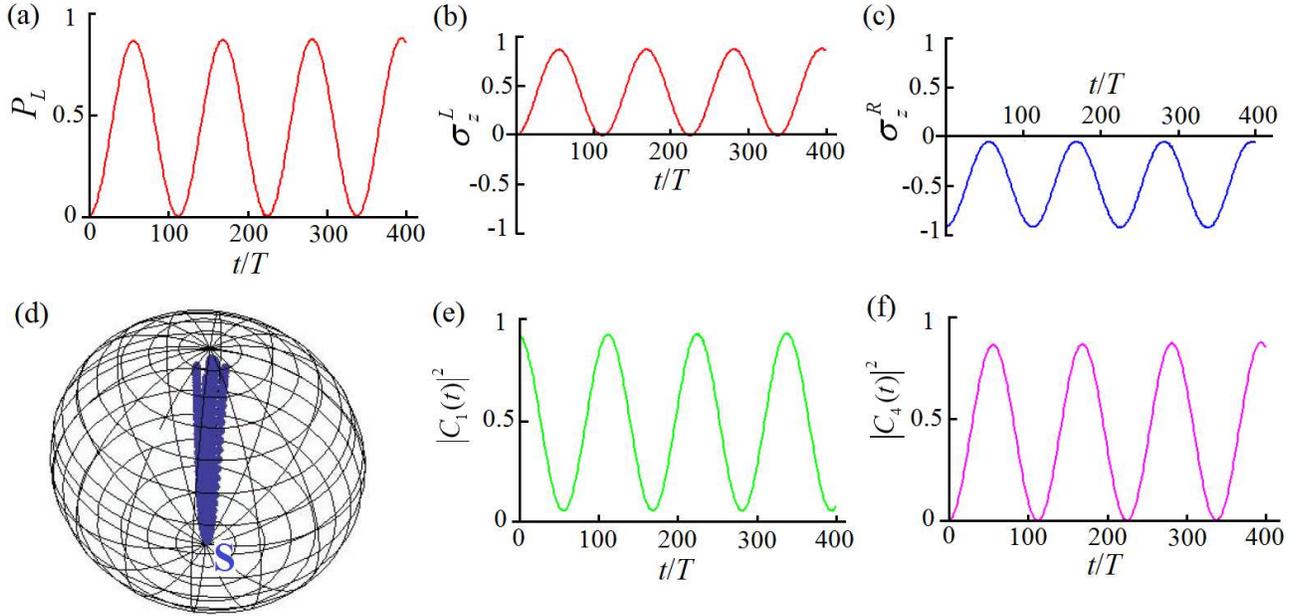}
\caption{(a)-(c) Same as in Fig.\ref{figa} shown for point B in Fig.\ref{figbzf}(b) for the spin-flip tunneling regime with (d) the spin vector dynamics within the Bloch sphere with the start point on the south pole S. Spin flip with the low-radius in-plane precession accompanying the tunneling is observed. (e), (f) Evolution of the occupation probabilities for the states $E_1$, $E_4$ participating in the dynamics.}
\label{figb}
\end{figure*}

Next we consider the point B in Fig.\ref{figbzf}(b) located on the angled spin-flip line (\ref{patflip}) with $k_2=6$. The evolution of observables and level population is shown in Fig.\ref{figb} in panels (a)-(c) and (e), (f), respectively. The tunneling probability period in panel (a) corresponds to the  tunneling time $\tau_f \sim 55 T$ which is in agreement with the ratio $\Delta_f/ \Delta_c \sim 0.45$ of the spin-flip $\Delta_f$ and the spin-conserving $\Delta_c$ tunneling matrix elements in our model so the spin-flip tunneling takes longer time as it can be seen by comparing Fig.\ref{figa}a and Fig.\ref{figb}a. The spin projection (\ref{sigmazl}) and (\ref{sigmazr}) in left and right QD demonstrate oscillations with the same period $2\tau_f$ as the tunneling probability but they have an opposite sign in left and right QDs. For the left QD one observes in Fig.\ref{figb}b that $\sigma_z^{L}(t)>0$ meaning that the spin is flipped during the tunneling and in the right QD one can see in Fig.\ref{figb}c that $\sigma_z^{R}(t)<0$ meaning that the spin is flipped back when the particle returns to the right QD. The corresponding spin-flip tunneling time is relatively fast, $\tau_f \sim 14$ ns, which is below the typical spin relaxation time in good quality GaAs samples and points to the possibility to observe such spin rotations experimentally.

Since the point B in Fig.\ref{figbzf}(b) represents an example of spin-flip dynamics in panel (d) of Fig.\ref{figb} we show the stroboscopic evolution of the the spin vector 

\begin{equation}
{\bf S}(t)=(\sigma_x(t),\sigma_y(t),\sigma_z(t))
\label{blochspin}
\end{equation}

shown within the Bloch sphere with the starting point at the south pole S. The mean values of all spin projections in (\ref{blochspin}) are defined as $\sigma_j(t)=\int_{-\infty}^{\infty} \langle \psi(x,t) |\sigma_j | \psi(x,t) \rangle dx$, $j=x,y,z$ where the area of the whole double dot system provides a contribution. This usual definition of the Bloch spin vector indicates certain differences with our plots of observables shown for a particular left or right QD. We thus consider the evolution of the Bloch vector (\ref{blochspin}) as an auxiliary tool indicating visually the regime of simple/complicated spin evolution. From  Fig.\ref{figb}d it can be seen that for the point B the spin demonstrates a flip along the z-axis accompanied with the slow and low-radius in-plane precession during the tunneling. It should be noted that in a multilevel system in the presence of SOI the spin is no longer conserved during the driven evolution and thus the spin vector (\ref{blochspin}) can be found not only on the surface but also inside the Bloch sphere \cite{Budagosky2016}. Finally, the evolution of the population of the states $E_1$ and $E_4$ is shown in panels (e) and (f) of Fig.\ref{figb} since these two levels are predominantly involved into the spin-flip tunneling corresponding to the spin-down state $E_1$ in right QD and the spin-up state $E_4$ in left QD. We can conclude that point B represents an example of predominantly two-level dynamics of tunneling between levels with opposite spins, which  can also be approximated by the analytical expressions (\ref{ptwolev}) and (\ref{patflip}).   

\begin{figure*}[tbp]
\centering
\includegraphics*[width=0.95\textwidth]{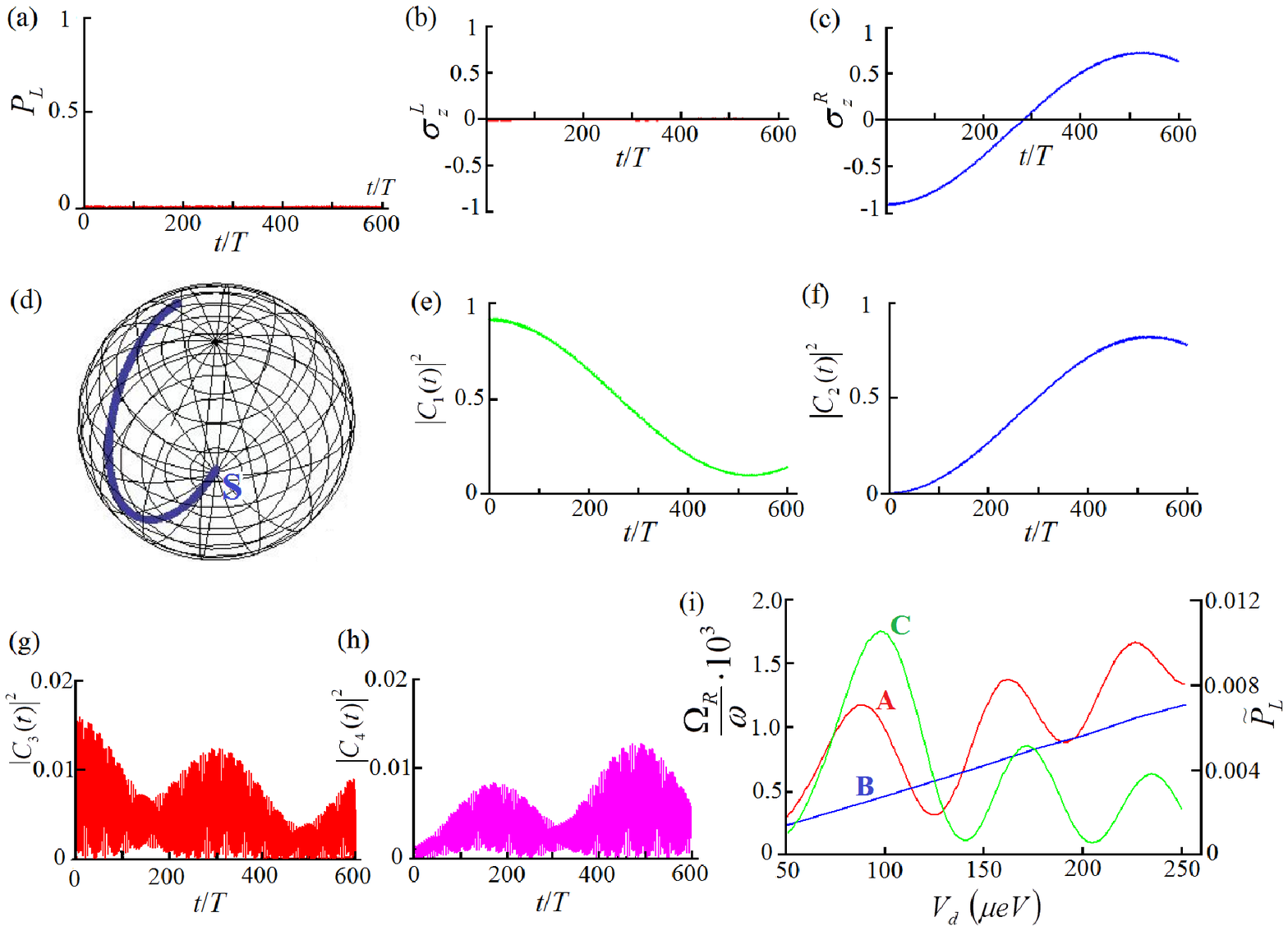}
\caption{Same as in Fig.\ref{figb} shown for point C in Fig.\ref{figbzf}(b) on 600 driving periods. The EDSR regime in right QD is observed with the spin flip (panel (c)) and without the effective tunneling (panels (a), (b)). The spin vector on Bloch sphere shown in panel (d) demonstrates a slow spin flip on $\sim 520$ driving periods starting from the south pole S and finishing near the north pole. (e), (f) Evolution of the occupation probabilities for the states $E_1$, $E_2$ from the right QD primarily participating in the dynamics and (g), (h) for the states $E_3$, $E_4$ in the left QD. (i) Dependence of the Rabi spin flip frequency $\Omega_R$ in units of the driving frequency $\omega$ on the driving amplitude $V_d$. The numerically obtained curve A for the four-level model is compared with the linear two-level dependence B. For curve A the spin flip is enhanced for the most of the driving strength range. Curve C is the tunneling probability ${\tilde P_L}$ (right axis) defined similar to (\ref{plr}) but averaged over one spin flip period and demonstrating the peaks coinciding with the ones for the Rabi frequency on curve A.}
\label{figc}
\end{figure*}

We move to the next point C in Fig.\ref{figbzf}(b) corresponding to the EDSR in the right QD and described by the angled line obeying (\ref{edsrcond}) with $k_3=1$, i.e. it is the basic EDSR line. The evolution of observables, the spin dynamics of the vector (\ref{blochspin}) on Bloch sphere and the level population dynamics are shown in Fig.\ref{figc} in the same sequence as in Fig.\ref{figb} but on the longer time interval of $600 T$. One can see that the tunneling probability in panel (a) and the spin projection in left QD in panel (b) are stable at almost zero value indicating that there is no effective tunneling in such regime. The spin projection in right QD shown in panel (c) demonstrates a slow spin flip with the flip time $\tau_f  \sim 520 T$ which is about $130$ ns. This slow spin flip can be seen in panel (d) for the spin vector (\ref{blochspin}) on Bloch sphere where the initial  point is on the south pole S and the end point can be observed near the north pole. The level occupations are shown in panels (e) - (h) where it can be seen that the two lowest states $E_1$ and $E_2$ in the right QD play the dominating part in the dynamics although the fast oscillating low-amplitude contributions from the states $E_3$ and $E_4$ are also present. 
For the two-level dynamics in the EDSR regime taking place in a single QD without the effective tunneling an estimate of the spin-flip Rabi frequency $\Omega_R$ was reported in Ref.\cite{Nowack2007}. In this paper the authors suggested the mechanism of EDSR due to the spatial oscillations of the QD potential minimum  by the applied electric field in the presence of SOI. Since these oscillations are significantly smaller in amplitude than the interdot oscillations (by approximately two orders of magnitude) the corresponding spin-flip Rabi frequency $\Omega_R$ is significantly smaller than the one observed for the spin-flip tunneling described above. 
In \cite{Nowack2007} the following estimate of the spin flip Rabi frequency was presented:

\begin{equation}
\Omega_R=\frac{g \mu_B |{\bf B}_{\rm{eff}}|}{2\hbar}
\label{rabinowack} 
\end{equation}

where the effective magnetic field created by SOI can be estimated as

\begin{equation}
|{\bf B}_{\rm{eff}}|=2B_z \frac{\Delta x_0}{l_{\rm{SO}}}.
\label{beffnowack} 
\end{equation}

In (\ref{beffnowack}) $B_z$ is the applied magnetic field, $l_{\rm{SO}}=\hbar^2/m\beta_D$ is the  spin-orbit precession length, and $\Delta x_0$ is the amplitude of the potential minimum displacement caused by the periodic electric field. The analytical solution for the potential minimum displacement can be derived from the explicit form of the sum of the double dot potential $U(x)$ and the detuning/driving potential $U_d(x)=U_d f_d(x)$ discussed in Sec.II. For the point C in Fig.\ref{figbzf}(b) we find that $\Delta x_0 \sim 0.2$ nm which is small compared to the spin-orbit precession length $l_{\rm{SO}} \sim 200$ nm or to the interdot travel distance $2d =116$ nm. We obtain from (\ref{beffnowack}) that $|{\bf B}_{\rm{eff}}| \sim 2 \cdot 10^{-3} B_z$ which according to (\ref{rabinowack}) gives the spin flip time $\tau_f \sim 500 T \sim 110$ ns for the point C in Fig.\ref{figbzf}(b). This estimate is close to the numerical result $\tau_f \sim 520 T$ seen in Fig.\ref{figc}c for this point. 

The description of the EDSR in a double dot system requires the discussion of a possible influence of the interdot tunneling on the spin flip. Spin-orbit interaction is more effective for longer traveling distances, therefore, we should take into account the tunneling between right and left QD since even small level populations in the left QD weakly coupled to the right QD can lead to noticeable spin flip events on longer than single tunneling time scales. The presence of even weak tunnel coupling to the other pair of spin levels in the neighboring QD visible in panels (g), (h) in Fig.\ref{figc} may produce certain differences to the EDSR mechanism during the driven level passage. It is known \cite{Khomitsky2012} that in a multilevel system the spin-flip Rabi frequency $\Omega_R$ can differ in its dependence on the driving amplitude $V_d$ from a simple two-level form $\Omega_R=V_{12}$ where $V_{12} \sim V_d$ is the matrix element of the driving field coupling the two spin states 1 and 2 participating in EDSR. The dependence on $V_d$ can become nonlinear if more than two levels participate in the dynamics. In Fig.\ref{figc}i curve A is the dependence of the Rabi spin flip frequency $\Omega_R$ in units of the driving frequency $\omega$ on the driving amplitude $V_d$ for our four-level model. Line B is the canonical two-level result for $\Omega_R$ discussed above. Note that all the panels in Fig.\ref{figc} except panel (i) are for the same point C in Fig.\ref{figbzf}b in the plane $(Bz, f)$ with fixed driving amplitude $V_d=100$ $\mu eV$ and in Fig.\ref{figc}i we move along the driving strength axis $V_d$ being perpendicular to this plane in the limits corresponding to the efficient tunneling. It can be seen that although the same growing trend with the increasing driving strength can be seen in both cases the numerically obtained data (curve A) differs from the linear two-level dependence B. 

The estimate of the effect of the tunneling on the spin flip frequency can be done as following: the contributions from the states $E_3$ and $E_4$ in the left QD compared to the ones in the right QD observed from Fig.\ref{figc}g,h give us the ratio $x_1 \sim |C_{3,4}|^2/|C_{1,2}|^2 \sim 10^{-2}$. This ratio is combined with the ratio $x_2$ of the interdot travel distance  $\Delta x_t \sim 2d$ and the potential minima displacement $\Delta x_0$ giving $x_2 \sim \Delta x_t/\Delta x_0 \sim 10^3$. The spin flip effectiveness can be expected as  proportional to both of $x_1$ and $x_2$ factors providing the variations of $\Omega_R/\omega$ to be of the order of unity which is observed in Fig.\ref{figc}i. By comparing curves A and B one can conclude that the spin flip frequency in the four-level case is mainly greater. The enhancement ratio of the spin flip frequency for four- and two-level models, is maximal for low and moderate driving fields which is favorable for practical applications. In Fig.\ref{figc}i another curve labeled as C is presented showing the tunneling probability ${\tilde P_L}$ (right axis) defined similar to (\ref{plr}) but averaged over one spin flip period. One can see that its peaks correlate well with the ones for the Rabi frequency on curve A reflecting the EDSR enhancement via the LZSM tunneling. The shape of the curve C resembles the one for the Bessel functions defining the tunneling probability in the two-level model (\ref{ptwolev}) in accordance with our results obtained for the tunneling regimes. We thus may call the discussed EDSR mechanism in Fig.\ref{figc} an LZSM-enhanced EDSR. It should be mentioned that a similar enhancement of the spin flip frequency has been reported recently for the EDSR modeling in a double quantum dot formed in silicon with the magnetic field gradient \cite{Burkard2022}. In that model both the interdot tunneling and the valley degree of freedom have been taken into account.

Another consequence of the LZSM fast level passage on the EDSR is the generation of several harmonics visible as lines 2, 3 in Fig.\ref{figbzf}b in addition to the main EDSR line 1 in Fig.\ref{figbzf}b. The higher harmonics generation can be viewed as a result of periodic sequence of short delta-like interaction pulses between the states in neighboring QDs during the periodic driving with a large amplitude leading to the short interaction time between the states.

\begin{figure*}[tbp]
\centering
\includegraphics*[width=0.95\textwidth]{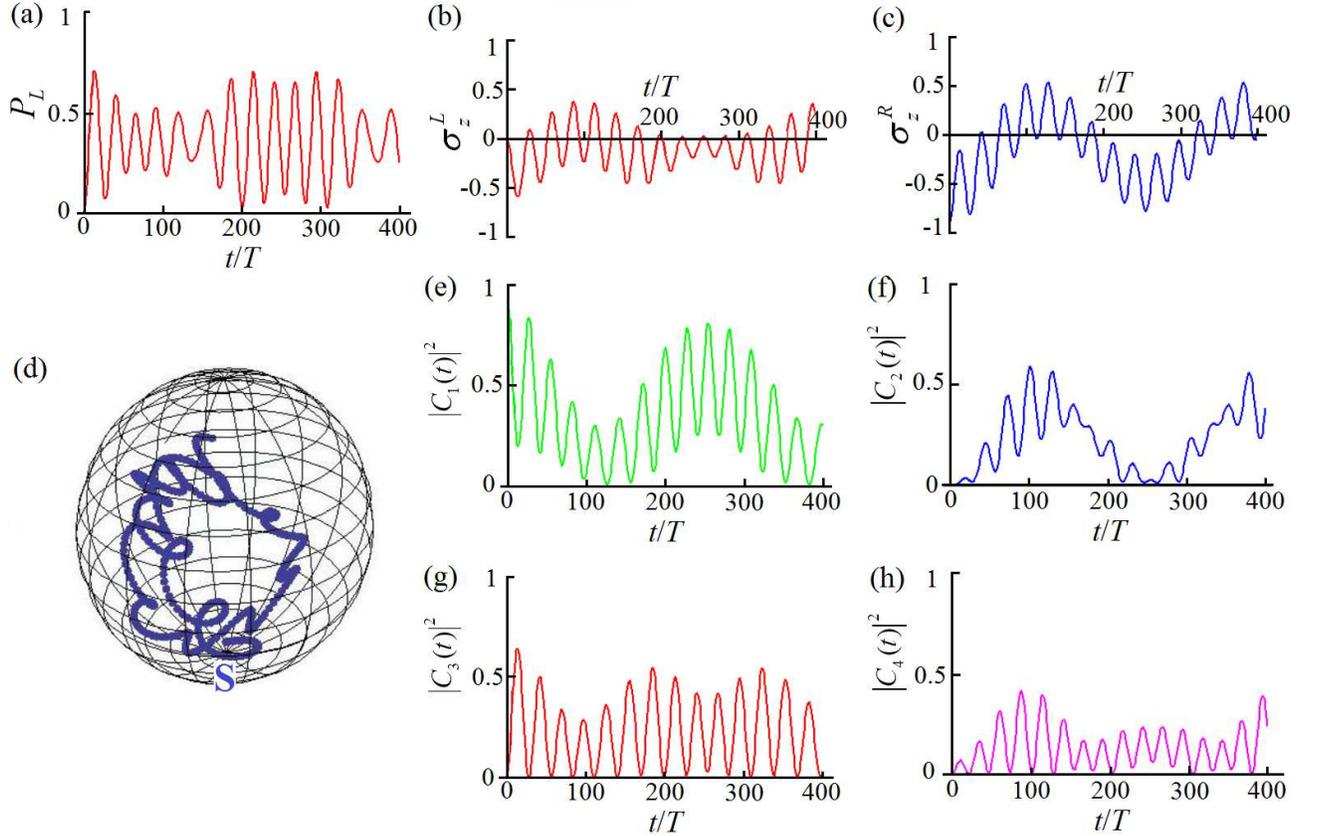}
\caption{Same as in Fig.\ref{figb} built for point D in Fig.\ref{figbzf}(b) showing the stroboscopic evolution in the hybrid resonance regime (\ref{hybrcond}) for (a) tunneling probability and (b), (c) spin projection in left and right QD demonstrating the tunneling with partial spin flip; (d) Evolution of the spin vector (\ref{blochspin}) within the Bloch sphere showing combination of spin flip and in-plane precession with the start point at the south pole S and the finish point near the north pole; (e) - (h) level occupations $|C_n(t)|^2$ demonstrating that all four states $E_1, \ldots, E_4$ provide comparable contributions  indicating that the system is in the multilevel regime far from the two-level approximation.}
\label{figd}
\end{figure*}

We conclude our analysis of the evolution for the selected points in the parameter space with the point D in Fig.\ref{figbzf}(b) corresponding to the hybrid resonance where all the conditions (\ref{pat}) - (\ref{edsrcond}) are satisfied together. The point D is characterized by the numbers $k_1=5$, $k_2=6$ and $k_3=1$ for the resonance conditions (\ref{pat}) - (\ref{edsrcond}) which satisfy the condition (\ref{hybrcond}) of the hybrid resonance resulting from the crossing of three types of resonances observed for the points A, B, C described above. The evolution of observables for point D is shown in Fig.\ref{figd}a - Fig.\ref{figd}c, the dynamics of the spin vector on the Bloch sphere is shown in Fig.\ref{figd}d, and the evolution of level occupations is shown in Fig.\ref{figd}e - Fig.\ref{figd}h. One can see that the tunneling probability dynamics in panel (a) demonstrates the tunneling on short times of about $20T$ typical for the spin-conserving tunneling under the condition (\ref{pat}). This tunneling is accompanied by the spin flip which happens according to the panels (b) and (c) in both left and right QD on a typical time scale $\tau_f \sim 100 T \sim 29$ ns which corresponds to the flip time for the spin-flip tunneling under the condition (\ref{patflip}) and is about four times faster than the spin flip under the EDSR condition in Fig.\ref{figc}. Overall, the spin and tunneling probability dynamics are modulated by a long-time envelope function with the period of around $300 T$ being typical for slow EDSR under the resonance condition (\ref{edsrcond}). We can conclude that the hybrid resonance have the traits of all three resonances found separately, namely, the spin-conserving tunneling, the spin-flip tunneling and the EDSR in a single dot. The spin dynamics on the Bloch sphere shown in panel (d) of Fig.\ref{figd} demonstrates a complicated behavior with traits typical both for short-radius rotations during the spin-flip tunneling (see Fig.\ref{figb}d) and the large-scale smooth dynamics typical for the EDSR in a single dot (see Fig.\ref{figc}d). The level population dynamics for point D shown in panels (e) - (h) in Fig.\ref{figd} clearly demonstrates that all the four levels $E_1, \ldots, E_4$ provide equal contributions to the dynamics. The basic frequency of the oscillations in Fig.\ref{figd}e - Fig.\ref{figd}h is the fastest frequency of spin-conserving tunneling. However, the spin can be flipped under such hybrid regime and not only in the left QD but also in the right QD as it can be seen in panels (b) and (c) of Fig.\ref{figd}, on the timescale which is significantly shorter than the one for the pure EDSR in a single dot presented in Fig.\ref{figc}. This finding can be important for design of the experiments and devices utilizing a fast spin control by purely electric driving fields.

If none of the conditions (\ref{pat}) - (\ref{edsrcond}) is satisfied then the system evolution corresponds in general to a dot placed in a dark background of the tunneling maps on Fig.\ref{FigVdrinvf}, Fig.\ref{FigVdrinvf2}b or in a light background in Fig.\ref{FigVdrinvf2}a shown above. In such point, as it is expected, no effective tunneling and no spin flip can be observed for the parameters considered in our examples for the double dot with a large detuning since the system rests in its initial state which is the ground state being the spin-down state $E_1$ in the right QD. 

To summarize this section, we see that the tunneling and the spin flip may manifest themselves in both separate and combined processes depending on the location in the parameter space. The effective tunneling can be realised in both spin-conserving (Fig.\ref{figa}) and spin-flip (Fig.\ref{figb}) regimes. In the latter case the spin flip takes place during tunneling to the neighboring QD. If the only EDSR condition is satisfied, a spin operation can be performed in the same QD where the spin is initialized (see Fig.\ref{figc}) but generally on a longer time scale due to the low contributions from the states of the neighboring QD and small amplitude of the potential minimum displacement in the individual QD. If one wishes a fast spin flip in the same QD where the spin has been initialized than a possible way to trigger it, according to our modeling, is by initializing a hybrid resonance regime. When the system parameters correspond to the condition (\ref{hybrcond}) then both the EDSR condition (\ref{edsrcond}), the spin-conserving (\ref{pat}) and the spin-flip condition (\ref{patflip}) are satisfied. Here we predict substantially faster spin flip taking place in both left and right QD with the spin flip time $\tau_{\rm{sf}} \sim 29 $ ns enhanced by SOI.
The final remarks that should be made in regards of the consideration of the noise and its influence on the spin decoherence time $\tau_2$. While a detailed study of noise can be a subject of a separate paper, one can estimate the spin decoherence time $\tau_2$ induced by the charge noise originating from the interdot tunneling as the main source of the charge noise since for the EDSR regime occurring mainly in a single dot the travelling distance is negligible. From the known models of charge noise in double dots with SOC \cite{Ludwig2014,Benito2019,Platero2022,Li2020} in the 100-mK temperature range which is relevant to the experiments \cite{Studenikin2018} that we are focused on these estimates give the range of $\tau_2$ in excess of 100 ns. The typical spin manipulation times obtained in the present modeling are within the 15...30 ns range indicating that one can execute at least several coherent spin rotations. 
Other mechanisms of spin decoherence can also be important and deserve further investigation which again is outside the scope of the present paper.

\section{Conclusions}

A multilevel LZSM-driven tunneling was studied theoretically under the conditions close to the recent LZSM experiments in a single hole DQD \cite{Studenikin2018}. The Floquet modeling of the driven dynamics revealed several remarkable features in the space of the system parameters. We carefully examine the situation when the condition for EDSR or its harmonics is satisfied. We predict the LZSM-enhanced EDSR and its harmonics that can be observed experimentally. 
The spin-dependent character of the tunneling is revealed in the 2D tunneling maps vs various system variables, i.e., the microwave frequency, driving amplitude, detuning, and the magnetic field. We explore the interference of the spin-conserved tunneling, the spin-flip tunneling, and the EDSR in a four-level Zeeman-split system in a DQD. We identify the conditions where  the three resonances mentioned above occur simultaneously.  In this condition of the hybrid resonance we predict the spin flip times being of around $50 \ldots 100$ periods for the driving frequency $f=2 \ldots 4$ GHz. This  gives us the the scale of $14 \ldots 29$ ns for both, the spin-flip during the tunneling and for the spin flip in a single QD. The efficiency of the spin-flip processes can be further optimised.  The results may be relevant for developing the schemes of spin control and readout in semiconductor devices by alternating electric fields.

\section*{Acknowledgements}

The authors are grateful to S. Ludwig, O.V. Ivakhnenko, R. Nath, M.V. Bastrakova, E.Ya. Sherman for stimulating discussions and to I.N. Budanova and Ya.S. Sergaev for technical assistance. D.V.K. is supported by the Ministry of Science and Higher Education of the Russian Federation through the State Assignment No 0729-2020-0058. 
S.A.S. acknowledges the support of the National Research Council Quantum Sensors Challenge Program.

\end{document}